\PassOptionsToPackage{table,xcdraw,dvipsnames}{xcolor}

\documentclass[preprint]{vgtc}         % preprint

\usepackage{microtype}                 % use micro-typography (slightly more compact, better to read)
\PassOptionsToPackage{warn}{textcomp}  % to address font issues with \textrightarrow
\usepackage{textcomp}                  % use better special symbols
\usepackage{mathptmx}                  % use matching math font
\usepackage{amsmath}
\usepackage{amstext}
\usepackage{amssymb}
\usepackage{times}                     % we use Times as the main font
\usepackage{cite}                      % needed to automatically sort the references
\usepackage{tabu}                      % only used for the table example
\usepackage{booktabs}                  % only used for the table example
\usepackage{wrapfig}
\usepackage{algorithm}
\usepackage{algpseudocode}
\usepackage{graphicx}
\usepackage{subcaption}
\usepackage{hyperref}
\usepackage{listings}
\usepackage[table]{xcolor}
\usepackage[normalem]{ulem}
\usepackage{float}
\usepackage{orcidlink}

\preprinttext{}

\title{A High-Performance SurfaceNets Discrete Isocontouring Algorithm}

\author{
Will Schroeder \orcidlink{0000-0003-3815-9386}
\thanks{e-mail: will.schroeder@kitware.com}\\%
\parbox{3in}{
\scriptsize \centering Scientific Computing, Kitware, Inc. 
}
\and
Spiros Tsalikis \orcidlink{0000-0001-5113-7195}
\thanks{e-mail: spiros.tsalikis@kitware.com}\\ %
\parbox{3in}{
\scriptsize \centering Scientific Computing, Kitware, Inc. \\ Computer Science, University of North Carolina at Chapel Hill
}\\
\and
Michael Halle  \orcidlink{0000-0002-0768-3233}
\thanks{e-mail: m@halle.us}\\%
\parbox{3in}{
\scriptsize \centering Brigham and Women\'s Hospital
}
\and
Sarah Frisken \orcidlink{0000-0001-5731-5095}
\thanks{e-mail: sfrisken@bwh.harvard.edu}\\%
\parbox{3in}{
\scriptsize \centering Brigham and Women\'s Hospital
}
}

\abstract{
Isocontouring is one of the most widely used visualization techniques. However, many popular contouring algorithms were created prior to the advent of ubiquitous parallel approaches, such as multi-core, shared memory computing systems. With increasing data sizes and computational loads, it is essential to reimagine such algorithms to leverage the increased computing capabilities available today. To this end we have redesigned the SurfaceNets algorithm, a powerful technique which is often employed to isocontour non-continuous, discrete, volumetric scalar fields such as segmentation label maps. Label maps are ubiquitous to medical computing and biological analysis, used in applications ranging from anatomical atlas creation to brain connectomics. This novel Parallel SurfaceNets algorithm has been redesigned using concepts from the high-performance Flying Edges continuous isocontouring algorrithm. It consists of two basic steps, surface extraction followed by constrained smoothing, parallelized over volume edges and employing a double-buffering smoothing approach to guarantee determinism. The algorithm can extract and smooth multiple segmented objects in a single execution, producing a polygonal (triangular/quadrilateral) mesh with points and polygons fully shared between neighboring objects. Performance is typically one to two orders of magnitude faster than the current sequential algorithms for discrete isosurface extraction on small core-count commodity CPU hardware. We demonstrate the effectiveness of the algorithm on five different datasets including human torso and brain atlases, mouse brain segmentation, and electron microscopy connectomics. The software is currently available under a permissive, open source license in the VTK visualization system.
}

\keywords{isocontouring, visualization, label maps, segmentation, atlases}

\begin{document}

%%
%% This command processes the author and affiliation and title
%% information and builds the first part of the formatted document.
\maketitle

\section{INTRODUCTION}

Increasing data size and the emergence of parallel computing systems remain powerful drivers in the current computing environment. However, many widely used algorithms were initially designed with a sequential computing model, meaning that their effectiveness in modern applications is greatly reduced. For example, Marching Cubes (MC) \cite{MC}, one of the most popular isocontouring algorithms \cite{VisHandbook}, employs computing patterns that are detrimental to high performance including multiple accesses to voxel values, repeated intersections of voxel edges, and use of centralized functions to insert output points and cells, and perform incremental memory allocation. Various approaches to improving performance have been proposed over the years, culminating with the recent introduction of the Flying Edges (FE) isocontouring algorithm \cite{FE}, which addresses many of the computational deficiencies of Marching Cubes, providing high-speed, scalable performance.

\setlength\intextsep{6pt}
\setlength{\columnsep}{9pt}%
\begin{wrapfigure}{R}{0.225\textwidth}
    \centering
    \includegraphics[width=0.225\textwidth]{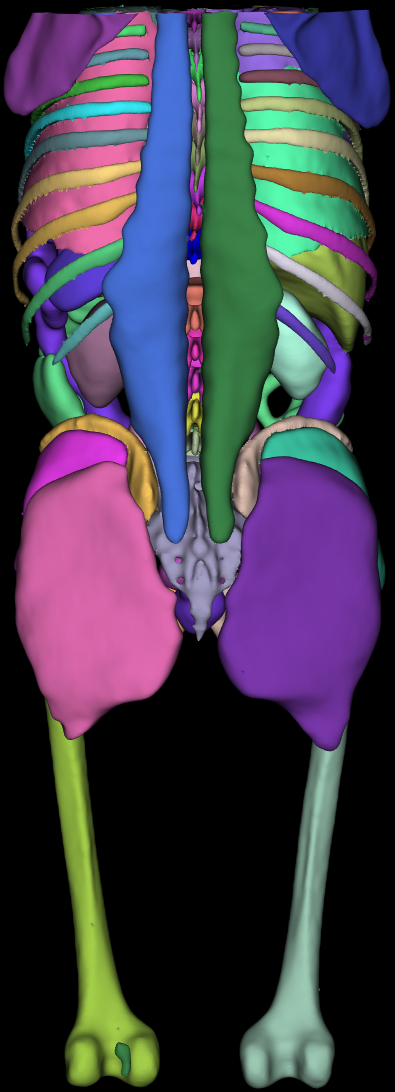}
    \caption{An AI-generated segmentation using TotalSegmentator \cite{TotalSegmentator} of a human torso. SurfaceNets extracted and smoothed 93 labeled objects. A random psuedo-color map is used.}
    \label{figure:torso}
\end{wrapfigure}

Another important driver in the computational environment is the increasing specialization and sophistication of scientific and engineering workflows. Again returning to Marching Cubes, which was initially designed to isocontour continuous scalar fields (continuous in the sense that scalar values are assumed to vary linearly across voxel edges), it was repurposed to contour discrete scalar field such as segmentation label maps. While this adaptation was certainly useful, as the demands on discrete isocontouring have grown, the MC adaptation has become increasingly problematic; for example, it produces surface meshes with serious deficiencies such as duplicated output polygons between adjacent labeled regions, so that boundaries between extracted objects are not fully shared. Further, to extract multiple isocontours, multiple passes are required for each contour value (i.e., for each segmentation label). Discrete Marching Cubes also produces “voxelized” surfaces which clearly reflect the resolution of the volumetric grid. Typically an additional smoothing pass is necessary to produce surfaces that better represent the actual object geometry.

\begin{figure}
    \centering
    \includegraphics[width=\linewidth]{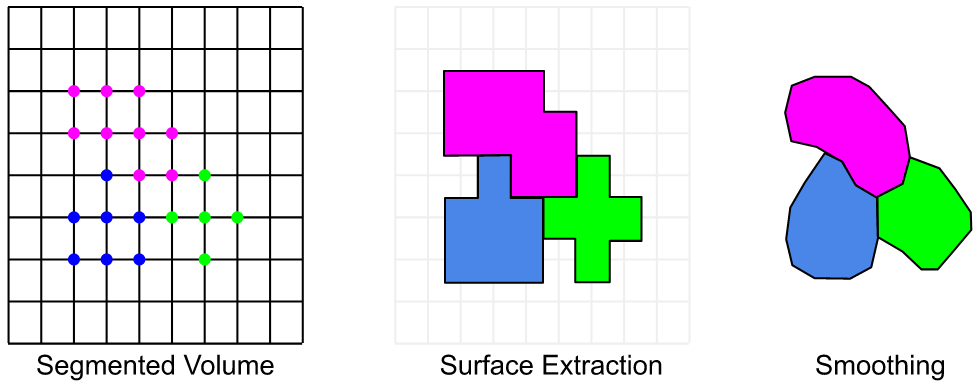}
    \caption{Overview of the SurfaceNets algorithm. A surface mesh encapsulating separate objects (i.e., segmented regions) is extracted from a labeled volume. Adjacent objects share boundary points and cells. The surface is then smoothed using a constrained Laplacian approach.}
    \label{figure:overview}
    \vspace{-10pt}
\end{figure}

SurfaceNets (SN) was developed in part to address the limitation of MC for discrete isocontouring, and to better support medical computing workflows such as anatomical atlas generation \cite{OpenAnatomy,AffordableAtlas}. As described by Frisken \cite{Frisken98,Frisken22}, at a high level the algorithm consists of two steps: 1) extraction of a voxelized surface, followed by 2) a smoothing of the surface (Fig.~\ref{figure:overview}). The surface extraction employs a dual contouring approach; that is, rather than generating points by intersecting voxel edges with an isocontour value, the extraction process generates at most one point at the center of a voxel. A point is generated when any one of its twelve edges intersects object boundaries, i.e., when the edge endpoints lie in separate labelled regions (including the background region). A quadrilateral polygon is also produced for each intersected voxel edge, connecting the voxel's center point to three other voxels sharing the intersected edge. At the same time, a smoothing stencil is defined consisting of up to six edges connecting voxel face neighbors which also contain a generated point. Note that all the isosurfaces (i.e., labeled regions) are simultaneously extracted in one execution of the algorithm. Once the surface is extracted, an iterative, constrained Laplacian smoothing process is performed using the stencils, in a process that constrains points to lie within the voxel from which they were generated. Typically after smoothing, the generated quadrilateral polygons are triangulated as the resulting surface is not planar. Also, as \cite{Frisken22} describes, the smoothing stencils are often modified to move points along boundary surfaces or edges, thereby improving the quality of the smoothing process.

In this paper, we present a novel, parallel, high-performance SurfaceNets algorithm. The algorithm borrows concepts from Flying Edges, including a multi-pass approach that processes data across volume edges, edge trimming to reduce computational load, and point and polygon generation without bottleneck functions such as incremental memory allocation and coincident point merging. In the following we describe the algorithm, offer important implementation details, and characterize performance on five representative datasets.

\section{ALGORITHM}

In this section we begin by providing a high-level overview of the SurfaceNets algorithm, including introducing descriptive notation. Key algorithmic features of the Parallel SurfaceNets (PSN) algorithm are then described.

\subsection{Notation} 
The SurfaceNets algorithm operates on structured grids with voxel scalar values arranged on a topologically regular lattice of data points (i.e., a data volume) with scalar values \(s_{i,j,k}\). Voxel cells \(v_{i,j,k}\) are defined from the eight adjacent points associated with scalar label values \(s_{i\pm{(0,1)},j\pm{(0,1)},k\pm{(0,1)}}\). Volume edges \(E_{i,j,k}\) run in the row \(E_{j,k}\), column \(E_{i,k}\), and stack \(E_{i,j}\) directions; and are composed of the union of voxel cell edges \(e_{i,j,k}\). For example the volume $x$-edges are defined by \(E_{j,k} = \bigcup_{i} e_{i,j,k}\). 

\setlength\intextsep{6pt}
\setlength{\columnsep}{9pt}%
\begin{wrapfigure}{R}{0.5\linewidth}
    \centering
    \includegraphics[width=\linewidth]{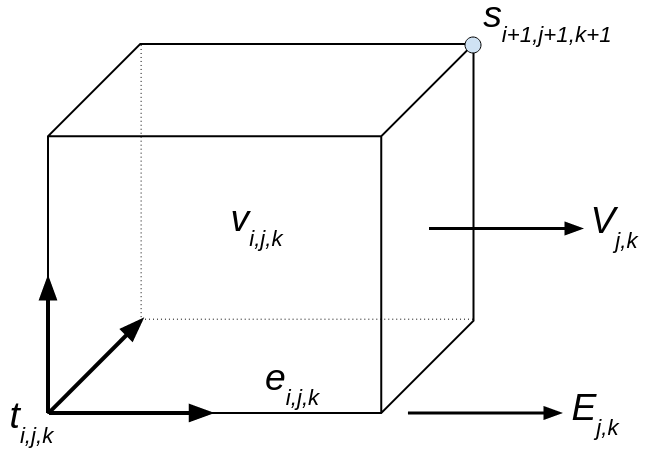}
    \caption{Depiction of parallel SurfaceNets concepts. The voxel cell $v_{i,j,k}$ with edge $e_{i,j,k}$ form volume rows $V_{j,k}$ and volume edges $E_{j,k}$. A voxel cell triad $t_{i,j,k}$ is associated with each $v_{i,j,k}$. Scalar label values $s_{i,j,k}$ are associated with each point in the volumetric lattice.}
    \vspace{-10pt}
    \label{figure:notation}
\end{wrapfigure}

The purpose of the algorithm is to generate an approximation to the set of isocontours \(Q_i\) with isovalues \(q_i\) such that \(q_i : Q_i(q_i) = \{p\, |\, F(p) = q_i\}\) where \(F(p)\) maps the point \(p \in R^N\) to a real-valued number. The resulting approximation surfaces \(S_i\) are continuous, piecewise linear surfaces such as a triangle mesh (in 3D). Note also that we say that \(Q_i\) intersects \(e_{i,j,k}\) when some $p$ lies along the voxel cell edge \(e_{i,j,k}\). 

This formulation presumes that multiple isosurfaces \(Q_i\) are simultaneously extracted, consistent with a volume defined by multiple discrete label values (i.e., a label map $L$, or segmentation mask). No presumption is made regarding the continuity of the scalar field; in fact in general the field is not continuous since the labels are typically disjoint. Consequently many isocontouring algorithms such as Marching Cubes which assume continuous (e.g., linear) variation across voxel edges, are not applicable to this problem, at least not without some assumption on continuity (e.g., voxel edges are intersected at their centerpoint). We presume that $L$ is a set consisting of multiple labeled regions \(l_i\), and \(s_{i,j,k} \in  L\), with the special definition of a background label $B$ consisting of any \(s_{i,j,k} \notin L\). Note to produce isocontours we choose some non-empty set \(q_i \in  L\). This produces labeled objects which approximate the isosurfaces \(Q_i\). 

Voxel cell rows \(V_{i,j,k}\) consist of all voxel cells \(v_{i,j,k}\) touching both \(E_{i,j,k}\) and \(E_{i+1,j+1,k+1}\). So for example, an x-row edge \(E_{j,k} = \bigcup_i e_{j,k}\) and \(V_{j,k} = \bigcup_i v_{i,j,k}\). Each \(v_{i,j,k}\) has an associated cell axes \(t_{i,j,k}\) (i.e., the voxel \emph{triad}) which is composed with the three cell edges emanating from the voxel origin located at \(s_{i,j,k}\) in the positive row $x$, column $y$, and stack $z$ directions. Refer to Fig.~\ref{figure:notation}.

\subsection{SurfaceNets}
SurfaceNets is an elegantly simple algorithm well suited to the extraction of labeled objects from volumetric objects. Conceptually, the twelve edges defining a voxel cell are intersected against the set $L$. An edge intersection is detected if for the two edge endpoints with scalar values \(s_0\) and \(s_1\), \(s_0 \neq s_1\) and \((s_0 \neq B\) or \(s_1 \neq B)\) (i.e., if \(s_0 = s_1 = B\), or \(s_0 = s_1\)  then no edge intersection occurs).

Once the edges of a voxel cell are intersected, a 12-bit edge case \(c^e_{i,j,k}\) can be generated to represent the intersection state (see Section~\ref{sec:facecase}). If any of the cell’s twelve edges is intersected, then a single point is generated in the center of the voxel cell. Further, for each intersected edge of \(c^e_{i,j,k}\) a quadrilateral polygon is generated orthogonal to and which bisects the edge. The quadrilateral connects the generated voxel cell center point with points located in the neighboring voxel cells sharing the intersected edge. Thus, depending on which edges are intersected, the generated point may be directly connected to up to six face neighboring cells, and up to 12 quadrilaterals may be generated (Fig.~\ref{figure:mcduplicates}). This connectivity defines the smoothing stencil which is the local network of edges connecting each generated point to its voxel face neighbors which must necessarily also contain a generated point. Note that the smoothing stencil is implicitly defined by the edge case: for every intersected edge, two voxel cell face neighbors are necessarily connected. Thus a face case \(c^f_{i,j,k}\) can be generated from the edge case and represented by a 6-bit value.

After processing all voxel cells, a quadrilateral surface mesh is generated. The mesh partitions the volume into regions (or objects) containing voxels with identical labels \(l_i\).  Typically this mesh is non-manifold since a voxel cell may be partitioned into eight separate regions with differing label values at each of its eight vertices. Note however that adjacent regions share boundary quadrilaterals and points, and each quadrilateral can be encoded with the label of the two regions on either side of it (a region adjacency two-tuple indicating the labeled regions on either side of it). At this point in the algorithm, the mesh appears “voxelized” consistent with its origins as a dual surface generation process. In the second major step of the SurfaceNets algorithm, the mesh is smoothed to remove sharp features and provide a pleasing visual result.

The smoothing process employs a modified Laplacian approach. For each point \(p_i\) its associated smoothing stencil is used to iteratively move \(p_i\) towards the averaged center point of its $N$ connected points $\Delta p_j$. A convergence factor $\lambda$ controls the amount of motion in a given iteration:
\begin{equation} p_i = p_i + \lambda \, \Delta p_j\text{,}\ \ \ \ \text{with}\ \ \ \Delta p_j = \frac{1}{N} \sum_{j=0}^N (p_j - p_i)\text{.} 
\label{eqn:smoothing}
\end{equation}

The total motion of \(p_i\) is constrained relative to its generating voxel cell. The constraints can be defined as a fraction of the hexahedral voxel cell, or more simply as a constraint sphere centered at the voxel center point. Finally, the smoothing stencils can be modified in such a way as to improve the smoothed mesh. For example, the motion of points along sharp edges can be constrained to the edge by judicious modification of the smoothing stencils (see \cite{Frisken22} for further details).

\subsection{Surface Extraction}
Highly-organized structured data such as a volume lends itself to a variety of parallel approaches. Our method takes advantage of cache locality by processing data in the fastest varying data direction (i.e., along the  volume $x$-edges); separating the processing into simple, independent computational passes along volume edges; reducing overall computation by performing a small amount of initial extra work (i.e., determine computational edge trimming); eliminating incremental memory allocation; and avoiding duplicate processing (such as determining voxel edge intersections multiple times on shared edges). 

The surface extraction portion of the algorithm is implemented as four passes across the data. The first three passes count the number of output points and triangles, bound the contour extent with computational edge trim values, and produce the metadata necessary to generate output in the fourth pass. In this last and final pass, output points, quadrilateral surface primitives, smoothing stencils, and the region adjacency two-tuple are generated and directly written into pre-allocated output arrays without the need for mutexes or locks. 

Note that the voxel triad \(t_{i,j,k}\) is a core concept to the algorithm described below. It carries five important bits of information about its associated voxel cell \(v_{i,j,k}\). These are: the classification of the scalar \(s_{i,j,k}\) (inside a labeled region, or background); whether the three $x,y,z$ voxel edges emanating from the cell origin intersect some \(Q_i\); and whether the associated voxel cell produces an output point. Triads can be computed independently in parallel and then combined to produce an edge case number as described previously and shown in Fig.~\ref{figure:edgecase}. Moreover, the triad encodes whether its $x-y-z$ axes should produce output quadrilaterals, and generate an output point. When combined, the generation of quadrilaterals from the triads eliminates multiple edge intersection operations, and produces an output mesh without duplicate faces.

In the following, we describe each of the four passes of surface extraction. While these passes are described in terms of processing $x$-edges, it is possible to process $y$ or $z$ edges instead. However, in typical data layouts, increasing $x$ corresponds to contiguous memory locations, thereby reducing cache misses.\vspace{0.05in}

\textbf{Pass 1: Process volume $\mathbf{x}$-edges \(\mathbf{E_{j,k}}\).}
For each volume $x$-edge \(E_{j,k}\), all voxel $x$-edges (i.e., \(e_{j,k}\)) composing \(E_{j,k}\) are visited and marked when their interval intersects \(Q_i\). The left and right trim positions \(x^L_{j,k}\) and \(x^R_{j,k}\) are noted, as well as the number of $x$-intersections along \(E_{j,k}\). Each \(E_{j,k}\) is processed independently and in parallel. (Note that the trim position \(x^L_{j,k}\) indicates where \(Q_i\) first intersects \(E_{j,k}\) on its left side; and \(x^R_{j,k}\) bounds where $Q$ last intersects \(E_{j,k}\) on its right side. Additional details describing computational edge trimming will be provided shortly.) Classification information about each triad is gathered (whether the voxel value \(s_{i,j.k} \in L\) or a background voxel \(s_{i,j.k} \in B\)).

\textbf{Pass 2: Process $y$- and $z$-voxel-cell edges.} For each volume cell row \(V_{j,k}\), the triads between the trim interval \([x^L_{j,k}, x^R_{j,k})\) are visited. The $y$- and $z$- triad edges are intersected against \(Q_i\). Depending on the outcome of this classification process, the trim interval may be adjusted to produce a final trim interval \([\bar{x}^L_{j,k}, \bar{x}^R_{j,k})\). Classification information gathered in Pass 1 is used to accelerate these intersection checks. For example, for a voxel edge with endpoints \((s_0,s_1)\), if \(s_0,s_1  \in B\) then no edge intersection exists.

\textbf{Pass 3: Configure output.}
Each voxel row  \(V_{j,k}\) is processed independently and in parallel. For each \(v_{i,j,k}\) in a row, the voxel edge case number \(c_{i,j,k}\) is computed by combining eight voxel triads forming the voxel cell (Section~\ref{sec:facecase}). If \(c_{i,j,k} = 0\), then no point is generated interior to the associated cell \(v_{i,j,k}\). Otherwise, a point and smoothing stencil are produced. If any of \(t_{i,j,k}\) edges are intersected, a quadrilateral polygon is produced. The number of output points, quadrilaterals, and smoothing stencil edges is recorded as per edge metadata. Once all \(V_{j,k}\) are processed, the metadata is updated via a prefix sum operation. This produces an indexing which maps the input data into output geometric primitives and stencils. To complete this pass, memory is precisely allocated for the output---no additional allocation is needed during the output generation process of Pass 4.

\textbf{Pass 4: Generate output.}
In the final pass, each \(V_{j,k}\) is processed in parallel by forward iteration (Section~\ref{sec:rowiterator}) over the \(t_{i,j,k}\) within the trim interval, producing center points and stencils in voxel cells as appropriate. Quadrilaterals are also output (if necessary) as each voxel triad indicates.

The trim interval plays an important role in the algorithm as it is used to rapidly skip data. Not only can the beginning and ending portions of \(V_{j,k}\) be skipped, but entire rows and slices of data as well. The triads \(t_{i,j,k}\) serve two important functions. First, they decouple computations into independent passes along volume edges. It is only when voxel cell information such as edge cases \(c^e_{i,j,k}\) are needed that are they combined using fast bit-wise operations. Secondly, the \(t_{i,j,k}\) control the generation of quadrilateral polygons. By combining the quadrilaterals generated from the eight \(t_{i,j,k}\) associated with the voxel cell vertices, the complete surface net is generated for each \(v_{i,j,k}\). Using the \(c_{i,j,k}\) and \(t_{i,j,k}\) to control primitive generation eliminates the need to merge coincident points and/or polygons, with points, quadrilaterals, and stencil edges generated only once and written directly into the previously allocated output arrays (Pass 3) without memory write contention.

\subsection{Smoothing} 
Once the surface has been extracted, and smoothing stencils defined, the second smoothing step of the Parallel SurfaceNets algorithm proceeds. To implement Equation~\ref{eqn:smoothing} in parallel requires approaches that avoid data races. While mutex or spin/speculative locks could be used, double-buffering avoids these slower mechanisms and ensures determinism. Basically, two arrays of points \(P_0\) and \(P_1\) are created. Initializing \(P_0\) with the output points of the surface extraction process, we used these points to compute \(P_1\). Next, we swap \(P_0\) and \(P_1\) and repeat (i.e., double buffer). Alternating swap / compute is used until the desired result is achieved (typically a modest number of iterations ~25 is used, although the algorithm is fast enough that the number of iterations and convergence factors can be adjusted interactively).

\subsection{Surface Triangulation}
Smoothing inevitably causes the mesh quadrilaterals to become non-planar. A final step in the algorithm is to triangulate the mesh. This can be trivially performed in parallel by generating two triangles from each quadrilateral. A variety of methods can be used to select the choice of tessellating diagonal: greedy, shortest diagonal, minimum area, and most co-planar are typical. The choice of triangulation method makes a modest impact on performance and appearance.

\section{IMPLEMENTATION DETAILS}

In the following section we highlight and discuss some of the implementation details of the Parallel SurfaceNets algorithm.

\subsection{Scalar Classification}
As described previously, the classification of the scalar \(s_{i,j,k}\) (inside a labeled region, or background) is retained by the voxel triad \(t_{i,j,k}\). In situations where the number of segmentation labels is large, scalar classification may significantly impact overall algorithm performance. We use specialized processes for classification as a function of the number of labels $N$ contained in $L$. A simple cache of last used label in $L$, and the most recent background label in $B$ greatly speeds classification. For a single label \(N=1\), a simple comparison with the cache returns set membership ($O(1)$ query time). For intermediate numbers of $L$ not in cache, label classification proceeds by comparison against a vector of $N$ values (search time $O(N)$ where $N$ is small. For large $N$, comparison is made against a set of values providing $O(\log(N))$ query complexity.

\subsection{Volume Metadata}
Volume metadata are used to cache intermediate results, limit the total amount of computation, and support the bookkeeping necessary to manage the generation of output. Our algorithm implementation requires two auxiliary metadata: the voxel triads, and the volume $x$-edge metadata. One voxel triad per voxel value is created (using an 8-bit unsigned char) which as described previously carry 5-bits of classification information about each voxel. To simplify implementation at the boundary, the volume of triads is padded out by one triad in each of the +/- $x$, $y$, and $z$ directions. Thus a volume of dimensions $M \times N \times O$, allocates a block of triad data of dimensions $(M+2) \times (N+2) \times (O+2)$. The volume $x$-edge metadata carries information about the primitives produced by each row of voxel cells \(V_{j,k}\): initially the number of points, the number of quadrilaterals, and the number of smoothing stencil edges is represented, but after the prefix sum in Pass 3 the volume $x$-edge metadata contains ids that control where the output is written in Pass 4. In addition, the metadata encodes the edge trim: the starting and ending voxels along \(V_{j,k}\) that contribute to the output. These five values are allocated for each padded volume $x$-edge using a 2D array of size $(N+2) \times (O+2)$.

\begin{wrapfigure}{R}{0.5\linewidth}
\centering
    \includegraphics[width=\linewidth]{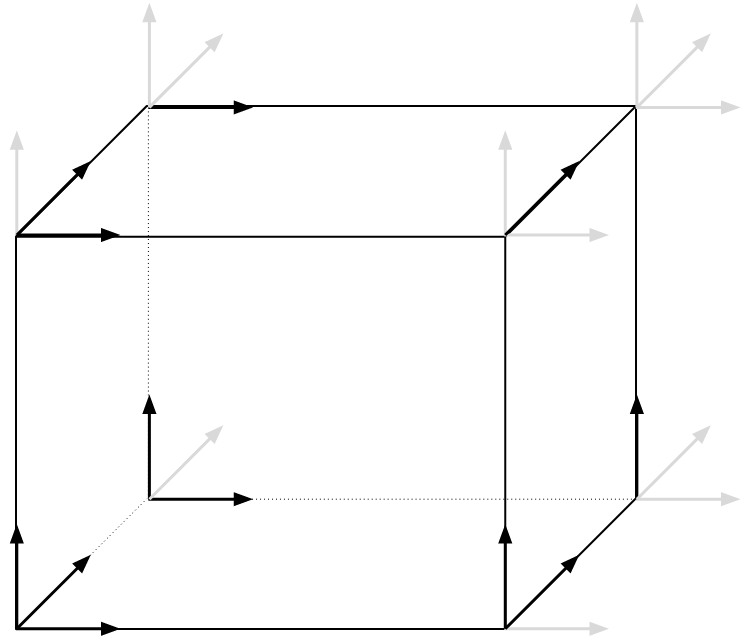}
\caption{The voxel cell edge case is determined by combining the eight voxel triads located at the eight vertices of the voxel cell. Only portions of the triads (bolded arrows) are used to determine the edges intersected by the labeled regions.}
\label{figure:edgecase}
\end{wrapfigure}

\subsection{Generating the Edge and Stencil Face Cases}
\label{sec:facecase}
The voxel cell edge case \(c^e_{i,j,k}\)can be easily computed from the voxel triads associated with eight voxel cell points (actually only seven triads are needed, and of those seven only portions of the triads). Recall that each triad carries information indicating whether the $x$, $y$, $z$ triad edges intersect some isosurface \(Q_i\). These data can be combined on the fly with simple bit-operations to generate a 12-bit edge case. Fig.~\ref{figure:edgecase} shows how portions of the triads are extracted and combined to produce the edge case. Similarly, a 6-bit stencil face case \(e^f_{i,j,k}\) can be computed from the edge case. For each intersected edge, the two voxel cell faces using that edge are selected. This selection process is repeated for each edge to set bits corresponding to the six faces of the voxel cell.

\subsection{Edge Trim}
Computational edge trimming is the process of pre-computing information or metadata in such a way as to reduce the need for subsequent computation, thereby reducing the total computational load. In the parallel SurfaceNets algorithm, computational trimming is used to significantly reduce the total effort necessary to extract the surface mesh. In typical application, large portions of the volume consist of background voxel values, especially towards the boundaries of the volume. Thus it is possible to rapidly skip portions of rows, entire rows, and even entire data slices while computing, significantly improving algorithm performance. As described previously, computational trimming is represented by the two tuple edge trim \([x^L,x^R)\) which are simply the left and right indices along each \(E_{j,k}\) which may produce output primitives (points, quadrilaterals, and stencils).

Computational trimming is determined over the course of the first two passes. In the first pass, evaluation of cell $x$-edges indicate the first and last voxels \(v_{i,j,k}\) on \(V_{j,k}\) which intersect the set \(Q_i\). These determine the initial edge trim \([x^L,x^R)\). In the second pass, the cell $y$- and $z$-edges are evaluated which may adjust the initial edge trim to produce the adjusted trim \([\bar{x}^L, \bar{x}^R)\). Note that there are certain pathological cases where the initial $x$-edge trim is empty, and contours intersect only the $y$- and/or $z$-cell edges. While the concept of edge trim is somewhat similar to run-length encoding, the approach described here only eliminates computation at the left and right side of the voxel edges \(E_{j,k}\) – interior runs of empty computation are not tracked. This approach is used to minimize memory overhead and performance impacts as compared to full run-length encoding, while still providing significant savings in many applications.

\subsection{Triad Output Point Generation}

\setlength\intextsep{6pt}
\setlength{\columnsep}{9pt}
\begin{wrapfigure}{r}{0.4\linewidth}
\centering
    \includegraphics[width=\linewidth]{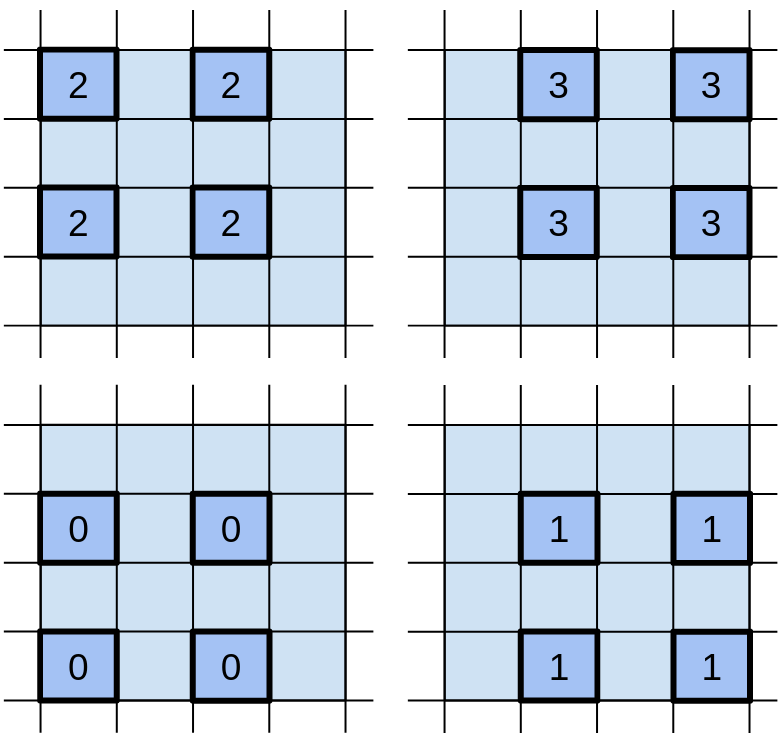}
\caption{In Pass 3, voxel cell rows $V_{j,k}$  (as viewed down the $x$-axis) are processed in a checkerboard pattern to avoid data races.}
\vspace{-10pt}
\label{figure:checkerboarding}
\end{wrapfigure}

In Pass 3 of the algorithm, the eight \(t_{i,j,k}\) of a voxel cell \(v_{i,j,k}\) are combined to produce an edge \(c^e_{i,j,k}\) and face \(c^f_{i,j,k}\) case. When \(c^e_{i,j,k} \ne 0\) then a point is to be generated in the center of \(v_{i,j,k}\). Recall that the \(t_{i,j,k}\) maintains a bit of information indicating whether a point is generated, so a non-zero \(c^f_{i,j,k}\) can be used to set this bit (the so-called \texttt{\small PRODUCE\_POINT} bit). However, during threaded processing, the \(t_{i,j,k}\) are being combined in neighboring voxel rows \(V_{i,j,k}\)---thus setting this bit introduces a potential race condition. There are several ways to address this situation including: 1) do not set the bit (i.e., recompute \(c^f_{i,j,k}\) whenever needed; 2) allocate a separate array to represent \texttt{\small PRODUCE\_POINT} bits; and 3) process voxel rows \(V_{j,k}\) in a coordinated fashion in such a way as to avoid data races. Through numerical experimentation, we determined that a \(2 \times 2\) checker boarding pattern of $E_{j,k}$ avoids data races with minimal impact on performance, and without the need to allocate additional memory. To perform the checker boarding, we group \(E_{j,k}\) into bundles of four neighbors (Fig.~\ref{figure:checkerboarding}), numbering them $0-3$ in each bundle. Then by processing all edges numbered $n$ in order (with four total threaded processing loops), race conditions are avoided since there is a guarantee that no \(t_{i,j,k}\) will be modified (i.e., its \texttt{\small PRODUCE\_POINT} bit set) during processing.

\subsection{Voxel Row Iterator}
\label{sec:rowiterator}
\setlength\intextsep{6pt}
\setlength{\columnsep}{9pt}%
\begin{wrapfigure}{R}{0.2\textwidth}
\centering
\includegraphics[width=0.2\textwidth]{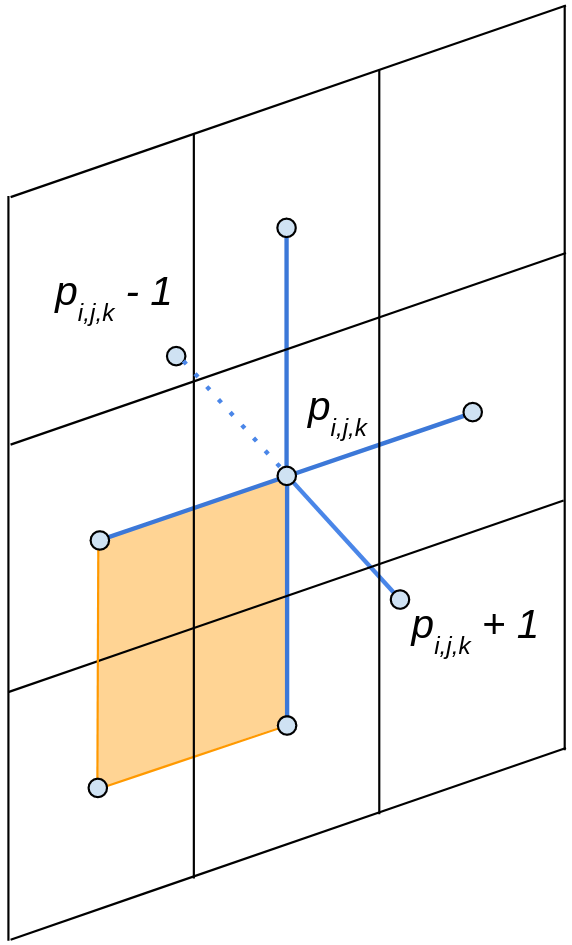}
\caption{Voxel row iterator for generating output points, quadrilaterals, and smoothing stencils. A potential triad $x$-edge quadrilateral is shown in orange.}
\label{figure:iterator}
\end{wrapfigure}

During the fourth and final pass, \(V_{j,k}\) are processed to produce output points, quadrilaterals, and smoothing stencils. This process is facilitated through the use of a voxel row edge iterator. The iterator is initialized with the point ids determined in the prefix sum operation in the third pass. Due to the smoothing stencils, which connect voxels in the current row to those in neighboring rows, a $3\times3$ iteration kernel is used to advance the point ids along \(V_{j,k}\) as well as its immediate voxel neighbors in the \(\pm j\) and \(\pm k\) directions (Fig.~\ref{figure:iterator}). These point ids are used to define a voxel center point, the (up to) three quadrilaterals generated by the \(t_{i,j,k}\), and the smoothing stencil edges. The performance of the row iterator is significantly improved through the use of the \texttt{\small PRODUCE\_POINT} bit in \(t_{i,j,k}\). As the iterator advances, each of the nine voxel triads in the iteration kernel are accessed and the \texttt{\small PRODUCE\_POINT} bit queried. If a bit value is set, then the point id is incremented. Subsequently when generating output, the iterator provides the appropriate point ids for producing output points, quadrilaterals, and stencils.

\subsection{Smoothing Cache}
The parallel SurfaceNets algorithm is often used in interactive workflows. Typically the surface extraction step is performed just once as the labeled objects to extract are known a priori. However, surface smoothing is often adjusted interactively to produce compelling results. This includes adjusting the number of smoothing iterations, and modifying the convergence factor $\lambda$. To facilitate this, our implementation caches the output of the surface extraction step, so that iterative smoothing workflows do not suffer from the cost of repeated surface generation.

\subsection{Load Balancing}
Many parallel isocontouring algorithms devote significant effort towards logically subdividing and load balancing the computation. Typical approaches include organizing volumes into rectangular sub-volumes or using a spatial structure such as an octree to manage the flow of execution. The challenge with isocontouring is that a priori it is not known through which portion of the volume the isosurface will pass, meaning that some form of pre-processing (evaluating min-max scalar region within sub-volumes for example) is required to balance the workload across computational threads. In the Parallel SurfaceNets algorithm, the basic task is processing of an edge, in which each edge (or associated voxel row) can be processed independently. In our implementation, we chose to use the Thread Building Blocks (TBB) library \cite{TBB} using the \texttt{parallel\_for} construct to loop in parallel over all edges. Behind the scenes, TBB manages a thread pool to process this list of edges, and as the workload across an edge may vary significantly, new edge tasks are stolen and assigned to the thread pool for further processing. Thus by designing the algorithm around appropriately sized computational worklets, task stealing is an effective approach to ensuring that processors remain busy.

\section{PERFORMANCE EVALUATION}
Algorithm performance was evaluated using workflows and a computing environment supportive of the anticipated user base. That is, the software was evaluated on a commodity laptop with an i9 processor with 10 physical cores (20 threads) and 64 GB of main memory, running Ubuntu Linux. While the algorithm was implemented as an MPI-capable filter in VTK, empirical evidence suggests that the majority of applications using PSN would run on conventional CPU desktop/laptop systems. Thus the evaluation shows what is possible on typical hardware, as well as the limitation of this computing environment when processing large volumes and/or numbers of labels (e.g., the connectome dataset as described in the following).

\subsection{Methods}
The performance of the Parallel SurfaceNets algorithm is evaluated against five datasets, and compared with three other algorithmic workflows. A summary of these five datasets follows.

\begin{figure*}[t]
\begin{center}
   \includegraphics[width=0.49\linewidth]{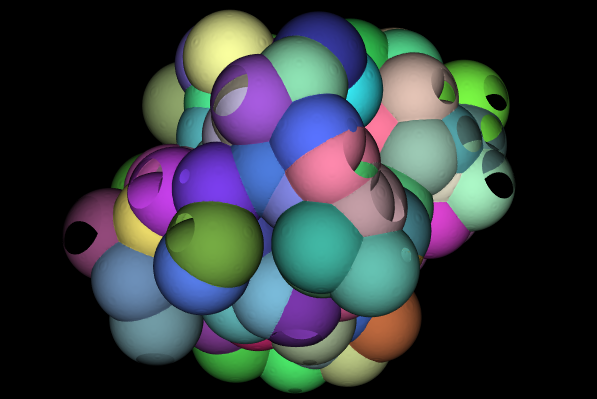}
   \includegraphics[width=0.49\linewidth]{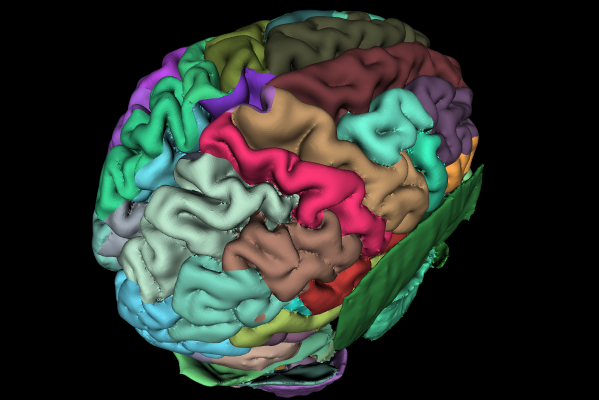} \\
   \includegraphics[width=0.49\linewidth]{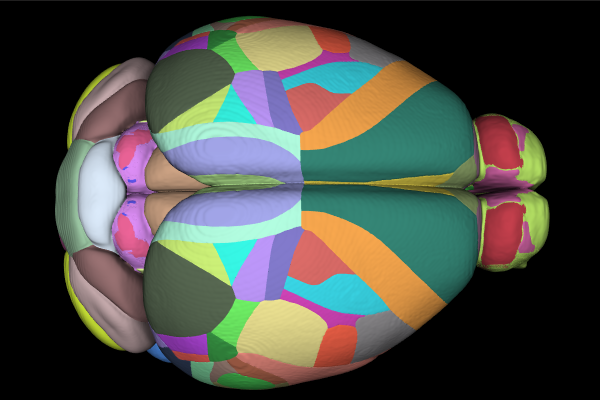}
   \includegraphics[width=0.49\linewidth]{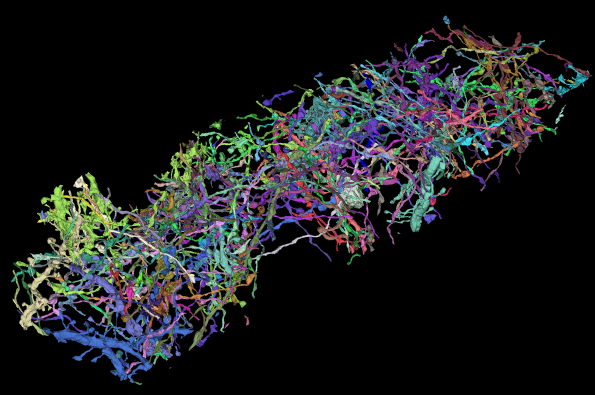}
\caption{Four of the five datasets used for performance testing. In reading order from the upper left: the Synthetic, Brain, Mouse, and Connectome datasets. Random colors are used for each segmented region. Fig.~\ref{figure:torso} shows the fifth Torso dataset.}
\label{figure:datasets}
\end{center}
\end{figure*}

\begin{description}
\item[Synthetic] a $512^3$ unsigned short volume with a random distribution of labeled, overlapping spheres. The number of spheres is 128, to bring the total number of labels to 129 (including the background label).
\item[Brain] is a segmented brain atlas of resolution $256^3$ with 313 labels. The data are a MRI-derived atlas of a normal volunteer provided by the SPL/NAC Brain Atlas \cite{BrainAtlas}.
\item[Torso] is a segmented human torso, segmentation performed by the AI-based TotalSegmentator \cite{TotalSegmentator}. The volumetric data has dimensions $317^2$ by 835 slices, with 93 segmentation labels. The data was segmented from 1204 CT images obtained during routine clinical practice \cite{Torso}.
\item[Mouse] brain atlas: Allen Mouse Brain Common Coordinate Framework (CCFv3) is a 3D reference space of an average mouse brain at 10um voxel resolution, created by serial two-photon tomography images of 1,675 young adult C57Bl6/J mice. The volume is of dimensions (1320,800,1140) with 688 labeled regions \cite{CellLocator}. 
\item[Connectome] is the IARPA MICrONS Pinky100 segmented connectome of the visual cortex of a mouse preoduced with high-resolution electron microscopy, segmentation, and morphological reconstruction of cortical circuits \cite{Pinky}. The test dataset is an extracted subvolume of dimensions $2048^3$ with 37,167 segmentation labels. Because of memory limitations in the MC algorithm (as described in the Section~\ref{sec:Discussion}), only 240 labels were extracted.
\end{description}
\noindent Selected views of the five datasets and their associated SurfaceNets extraction is shown in Fig.~\ref{figure:datasets}.

Using these five datasets, four different algorithmic workflows performing discrete isocontouring of annotated segmentation are compared. While each workflow varies in the particulars of its application and output, in general they produce similar results and have been used in practice (for example in 3D Slicer \cite{Slicer}). The first workflow is based on a modified version of Marching Cubes for discrete data (referred to as the MC workflow). In this version, because the data is discrete and non-continuous, edge intersections are forced to be at the center of voxel edges whose end points are in different labeled regions. Each segmentation label results in a separate contouring operation. Generated points are shared across adjacent labeled region boundaries, although output triangles are not (resulting in topological and geometric gaps, and producing duplicate triangles Fig.~\ref{figure:mcduplicates}). In this first workflow, once the output triangle surface is generated, an iterative smoothing pass with 25 iterations is initiated using Taubin's \cite{TAUBIN} windowed sinc fairing algorithm. MC is a sequential workflow, so comparisons are made using one thread of execution.

\begin{table*}[h]
\begin{center}
    \begin{tabular}{|l|r|r|r|r|r|r|r|r|r|r|}
    \hline
   {\bf Algorithm} & \multicolumn{2}{c|}{\bf Synthetic} & \multicolumn{2}{c|}{\bf Brain} & \multicolumn{2}{c|}{\bf Torso} & \multicolumn{2}{c|}{\bf Mouse} & \multicolumn{2}{c|}{\bf Connectome} \\
    \cline{2-11}
    & Time (s) & Speed Up & Time (s) & Speed Up & Time (s) & Speed Up & Time (s) & Speed Up & Time (s) & Speed Up \\
    \hline
    MC & 51.000 & 1.00 & 4.5851 & 1.00 & 4.7540 & 1.00 & 1421.00 & 1.00 & 378.60 & 1.00 \\
    FE (1 thread) & 22.160 & 2.30 & 5.9690 & 0.77 & 8.0830 & 0.59 & 768.10 & 1.85 & 1730.00 & 0.22 \\
    FE (16) & 2.760 & 18.48 & 0.8225 & 5.57 & 1.0360 & 4.59 & 103.60 & 13.71 & 222.70 & 1.70 \\
    MM & 31.230 & 1.633 & 5.0330 & 0.91 & 14.2200 & 0.33 & --- & --- & --- & ---\\
    \hline
    PSN (1 thread) & 3.420 & 14.89 & 0.5546 & 8.27 & 0.8653 & 5.59 & 33.35 & 42.59 & 74.87 & 5.06 \\
    PSN (2) & 1.760 & 28.97 & 0.2804 & 16.34 & 0.4526 & 10.50 & 17.18 & 82.70 & 38.86 & 9.74 \\
    PSN (4) & 0.951 & 53.62 & 0.1435 & 31.94 & 0.2484 & 19.13 & 9.191 & 154.50 & 20.98 & 18.04\\
    PSN (8) & 0.573 & 89.00 & 0.0838 & 54.72 & 0.1577 & 30.14 & 5.860 & 242.40 & 13.00 & 29.13 \\
    PSN (16) & 0.405 & 125.81 & 0.0561 & 81.74 & 0.1168 & 40.70 & 4.552 & 312.00 & 9.98 & 37.95 \\
    \hline
     \#pts / \#tris & \multicolumn{2}{c|}{\small 5,213,497 / 10,551,274 } & \multicolumn{2}{c|}{\small 1,019,951 / 2,155,766} & \multicolumn{2}{c|}{\small 1,405,615 / 2,867,390} & \multicolumn{2}{c|}{\small 54,568,063 / 111,823,070} & \multicolumn{2}{c|}{\small 66,568,012 / 133,275,812} \\
    \hline
    \end{tabular}
\end{center}
\vspace{0.1in}
\caption{Comparison of algorithm performance across five different datasets. The sequential discrete Marching Cubes (MC) and parallel discrete Flying Edges (FE) algorithms are shown along with the sequential Multi-Material SurfaceNets reference algorithm (MM) available from \cite{Frisken22}. The Parallel SurfaceNets algorithm proposed here is shown with varying numbers of computational threads. Along with times of execution, speed up factors (based on MC performance) are also tabulated. Performance is measured on a commodity laptop (i9, 10 cores, 64 GB memory, Ubuntu Linux). The final row shows the number of points and triangles generated by PSN for each dataset. Note that MM was unable to process the larger datasets due to memory constraints so results are not shown for these cases.}
\label{table:performance}
\end{table*}

The second workflow, similar to the first, is based on a modified discrete Flying Edges algorithm (FE). Again, intersections are forced at edge centers. Each segmentation label results in a completely independent execution of FE, producing duplicate points and triangles. The windowed sinc smoothing algorithm is also used to produce the final smoothed results. Note that in both the first and second MC and FE workflows, smoothing can cause the surfaces of adjacent objects to intersect or pull apart because boundaries are not fully shared. FE has been designed for threaded computation, so that comparisons are made for 1 and 16 threads of execution. 

The third and fourth workflows are based on SurfaceNets. The third workflow is the Multi-Material (MM) algorithm described by Frisken \cite{Frisken22} . The author provided the source code to enable direct comparison with the other workflows. MM is a sequential algorithm. Finally, the fourth workflow is the Parallel SurfaceNets algorithm described in this paper. Because PSN is threaded, comparisons are made using 1, 2, 4, 8, and 16 threads of execution. Note that in both MM and PSN, the SurfaceNets constrained smoothing process is invoked with 25 iterations.

To measure performance, an initial run was performed to preload data and the executable program into computer memory. Then ten runs capturing elapsed time were averaged to produce a final execution time. The run times and speed up factors are shown for the four workflows (MC, FE, MM, and PSN) described above, across the five test datasets with (where applicable) variable thread counts. The number of points and triangles produced by PSN is shown at the bottom of the table.

\begin{figure}[b]
\centering
   \includegraphics[width=1.0\linewidth]{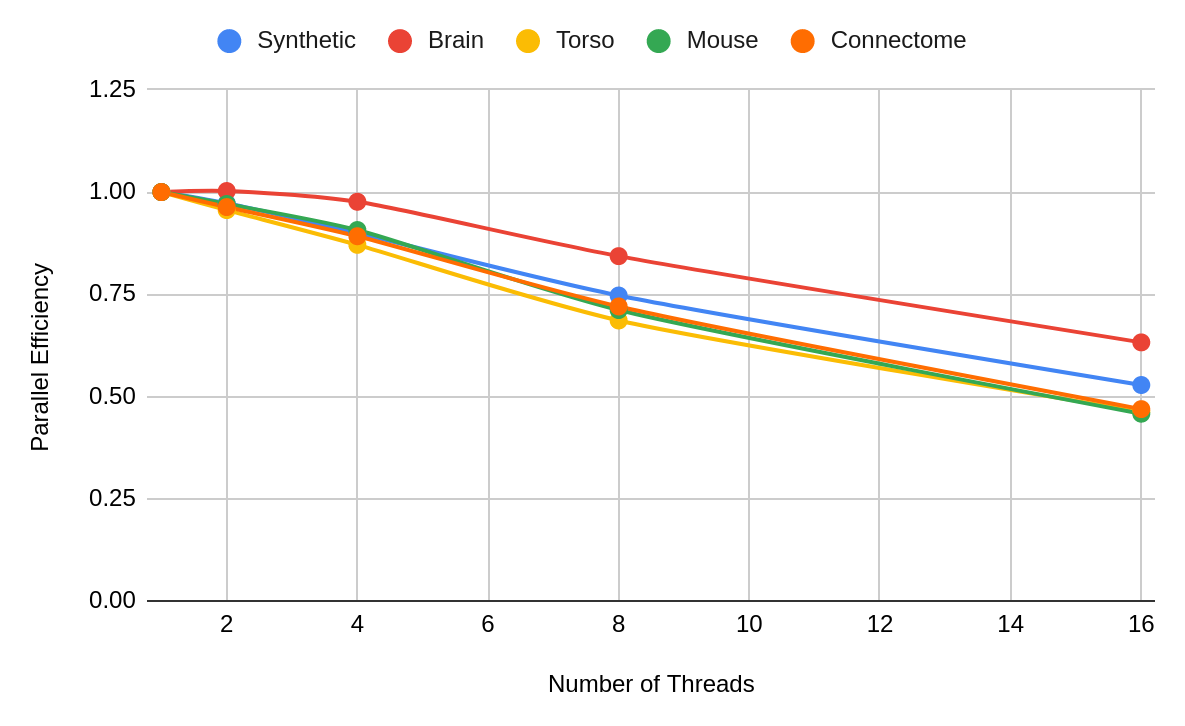}
\caption{Parallel efficiency curves from Table~\ref{table:performance}.}
\label{figure:efficiency}
\end{figure}

Parallel efficiency curves for the PSN algorithm is shown for the five datasets in Fig.~\ref{figure:efficiency}. Finally, the effect of the number of extracted labeled objects is shown in Fig.~\ref{figure:slowdown}. To vary the number of labeled objects, we used the Synthetic dataset and simply modified the list of labels to produce the desired number of extracted regions. 

\begin{table*}[h]
\centering
    \begin{tabular}{|l|r|r|r|r|r|r|}
    \hline
    {\bf DataSet} & \multicolumn{4}{c|}{\bf Surface Extraction} & {\bf Smoothing} & {\bf Triangulation} \\
    \cline{2-5}
    & Pass1 & Pass2 & Pass3 & Pass4 &  &  \\
    \hline
    Synthetic & 9.78\% & 11.55\% & 16.98\% & 29.12\% & 18.13\% & 14.44\% \\
    Brain & 13.70\%	& 5.91\%	& 7.41\%	& 29.72\%	& 23.78\%	& 19.47\% \\
    Torso & 19.97\%	& 7.48\% & 12.14\% & 24.51\% & 19.92\% & 15.99\% \\
    Mouse & 8.94\% & 9.21\%	& 12.32\% & 29.68\% & 22.54\% & 17.30\% \\
    Connectome & 12.37\% & 15.24\% & 26.44\% & 33.08\% & 6.97\% & 5.91\% \\
    \hline
    \end{tabular}
\vspace{0.1in}
\caption{Percentage of the time spent in the major portions of the Parallel SurfaceNets algorithm. The four passes correspond to the surface extraction process to produce quadrilateral polygons; this is followed by smoothing and then trianglution of the quadrilaterals. Results shown are for 16 threads and 25 smoothing iterations for the five test datasets.}
\label{table:breakdown}
\end{table*}

Table~\ref{table:breakdown} shows the time to execute the major steps of the PSN algorithm for each of the five test datasets. The time is expressed as a percentage of the total algorithm execution time, computed with 16 threads. The steps shown are the four passes of the surface extraction process, followed by smoothing and final triangulation.

\subsection{Discussion}
\label{sec:Discussion}
Table~\ref{table:performance} clearly captures the speed improvements of PSN versus the other workflows. Even executing with just a single thread the algorithm is faster than the other three. When threading is enabled, speedups range from one to two orders of magnitude topping out at 312$\times$ faster for the Allen Mouse Brain Atlas. Parallel efficiency for PSN falls rapidly from 8 to 16 threads. Based on empirical evidence, we believe that this effect is due to exceeding the number of physical CPU cores (10), plus the high rate at which output points, triangles, smoothing stencils, and cell neighbor two-tuples are generated, negatively impacting memory bus performance. In future work we will further examine efficiency impacts to identify potential performance improvements.

As evidenced by the Connectome dataset results, the Parallel SurfaceNets algorithm can generate large meshes very quickly (i.e., over 130 million triangles in under 10 seconds for 240 labels). While processing the entire dataset, PSN successfully completed processing all 37,167 labels in the first three passes. However, in Pass 4 an attempt was made to allocate memory for billions of points, quadrilaterals, and stencils. This request, when added to the $2048^3$ labeled volume and algorithm metadata, exceeded the 64 GB computer main memory. Fortunately PSN enables extraction of subsets of the labeled regions. The other workflows (MC,FE,MM) also failed on this dataset, with the MC reference algorithm topping out at 240 labels.

\begin{figure}[t]
\centering
   \includegraphics[width=1.0\linewidth]{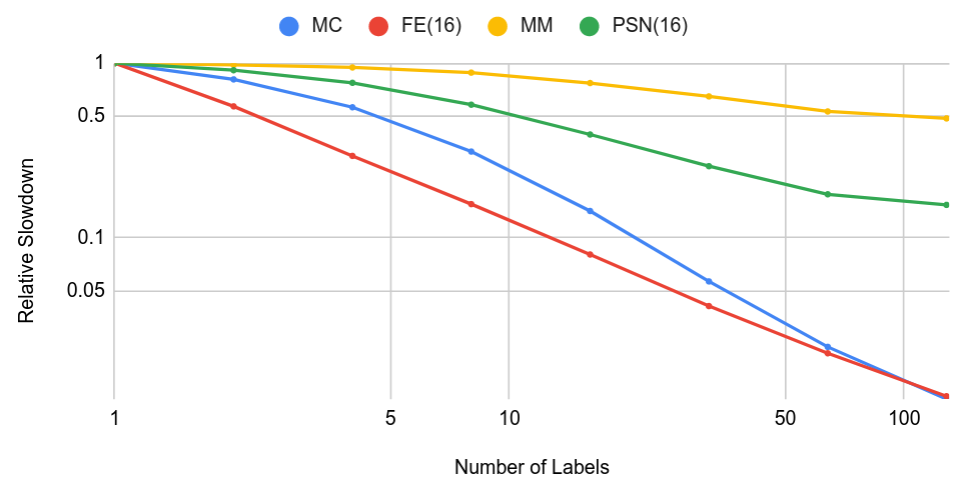}
\caption{Relative slowdown as the number of labels increases. Curves produced with the Synthetic dataset for the number of labels ranging from [1,128] and plotted with logarithmic scale.}
\label{figure:slowdown}
\end{figure}

MC and FE also exhibit some interesting behavior. Because FE executes a separate processing pass for each label value, its performance is significantly impacted as the number of labels increases, since each processing pass requires a complete traversal of the input volume---this is apparent from the linear behavior of FE performance as the number of labels increases Fig.~\ref{figure:slowdown}. In fact, if the number of labels is reduced to small numbers, FE eventually becomes faster than PSN. This is because FE uses geometric reasoning to extract manifold isosurfaces; an advantage over Parallel SurfaceNets as PSN assumes non-manifold surfaces since multiple materials (up to eight) can join in a single voxel cell. Also note that the curves for PSN and MM in Fig.~\ref{figure:slowdown} do not track closely as might be expected. As implemented, MM performs an expensive initialization process to classify every edge in the volume, which tends to flatten its slowdown curve. In comparison, PSN uses computational trimming, so that many $y$- and $z$-voxel edges do not require explicit classification. Thus the impact on PSN performance as the number of labels increase is closely related to processing those $v_{i,j,k}$ that actually produce mesh points and polygons.

MC performs surprisingly well for smaller datasets despite suffering from several parallel deficiencies including incremental memory allocation as points and cells are inserted, and the use of a spatial point locator for merging duplicate points. Internally the discrete MC version traverses the volume just once, paying the cost of loading a voxel just one time, and then for each voxel quickly looping over all contour values to generate output. Because duplicate points occur only at edge midpoints, the demands on the locator are reduced as fewer points are generated across voxel edges (a maximum of 12 points per voxel). However, for larger volumes the MC binning algorithm can become overwhelmed and performance suffers badly (which is clear in the Mouse dataset results shown in Table~\ref{table:performance}). 

\begin{figure}[t]
\centering
   \includegraphics[width=1.0\linewidth]{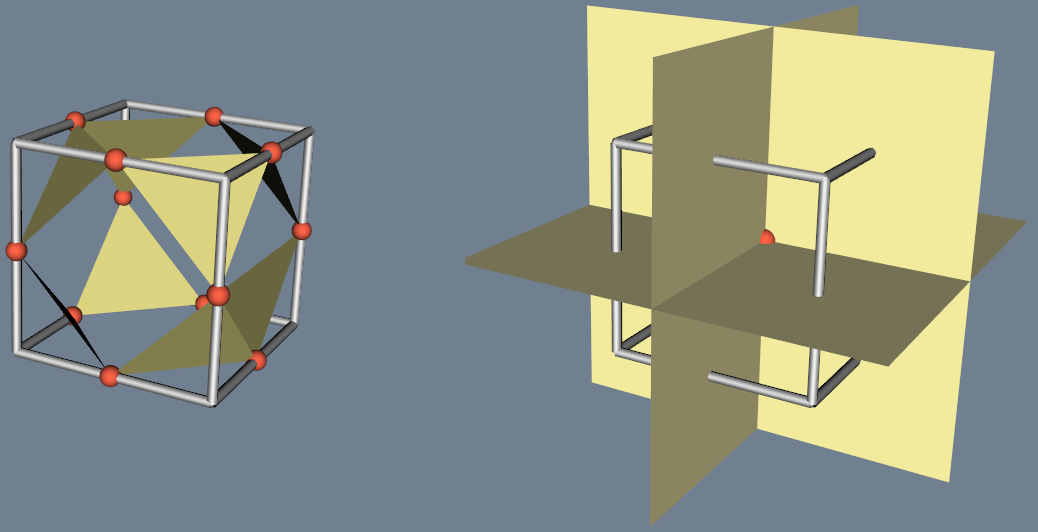}
\caption{A comparison of voxel cell subdivision when all eight cell vertices lie in different regions. Discrete MC (left) generates 12 shared points (in red), and eight duplicated triangles. SurfaceNets generates a single interior point (in red), and 12 shared quads connecting neighboring voxels.}
\label{figure:mcduplicates}
\end{figure}

Beyond speed, Parallel Surface Needs produces output meshes without duplicate points or triangles. FE produces both duplicate points and triangles on shared region boundaries. While this can be repaired with a post-processing step, given the significant speed advantage of PSN there is little motivation to do so. While MC eliminates duplicate points, it produces "gaps" between objects. For example, if the eight vertices of a voxel cell lie in different labeled regions, the Marching Cubes case table will create eight triangles that separates the eight vertices into separate regions (see Fig.~\ref{figure:mcduplicates}) leaving gaps between regions. Because FE and MC do not fully share the boundary between adjacent regions, this may cause difficulties in smoothing as described previously. Moreover, both FE and MC will produce output of a greater size due to duplicate mesh entities.  For this reason the MC algorithm limited the number of labels that could be extracted in the Connectome dataset to 240. This is due to MC's creation of duplicate faces, and the memory allocated to the spatial point locator.

Note that PSN and MM produce similar results, with differences found on the boundary (i.e., where segmented regions abut one or more of the six volume faces). The original MM implementation choose to \emph{cap} segmented regions, whereas the PSN implementation leaves segmented regions open. The reason for this is that, in very large datasets, PSN can be run in distributed fashion across multiple sub-volumes using for example MPI-based data parallel approaches. By not capping segmented regions, the resulting distributed execution avoids introducing artificial partitions along adjacent sub-volume boundaries.

Table~\ref{table:breakdown} clearly shows that they majority of time is spent generating output primitives in Pass 4 of the surface extraction process. As the volume increases in size, more total time is spend extracting the surface. Consequently smoothing and triangulation consume a larger percentage of the total work as the volume decreases in size.

\section{SUMMARY and FUTURE WORK}
We have developed a high-performance, scalable isocontouring algorithm for discrete, segmented scalar data. The Parallel SurfaceNets algorithm produces an output triangle mesh that encloses each segmented region. Based on evaluation across five different segmented datasets, sequential performance was greater than three other comparable workflows, while parallel performance on 16-threads with commodity hardware was consistently one to two order of magnitude faster. The PSN algorithm achieves its speed by performing independent threaded operations across volume edges, resulting in scalable performance through task-stealing, edge-based computational tasks. It eliminates sequential bottlenecks such as incremental memory insertion and centralized binning to remove duplicate points, while adopting a multi-pass approach combined with computational trimming to progressively eliminate processing in subsequent parallel passes. The algorithm produces an output mesh that fully shares points and triangles across neighboring objects. Moreover, triangles shared between objects can be annotated to indicate which objects are adjacent to each other, opening up the possibility of topological query and analysis.

Since large segmented volumes may exceed computer memory resources, distributed, hybrid parallel computing approaches may be required in future applications. Such hybrid approaches could combine local, shared-memory threading with distributed computing to process subvolumes from a larger input volume. As currently implemented, PSN can be employed in this manner, with special attention place on stitching output meshes together at subvolume boundaries. However, given the enormous size of meshes that can be generated by PSN in datasets such as brain connectomics, another approach may be to generate output meshes on demand as part of an interactive visualization process or data query. PSN is fast enough that subvolumes can be dynamically processed as they become visible during interaction, or in response to requests to view certain structures. Other avenues of investigation include on-the-fly decimation \cite{Decimation} of output polygon streams to reduce the size of the output.

An important feature of SurfaceNets is the generation of the two-tuple region adjacency information. By combining this with annotation ontology and classification, it us possible to perform topological queries such as locating objects adjacent to other structures, e.g., locate all organs adjacent to a vascular structure. This capability if of particular importance in medical applications where anatomical atlases are under construction for educational, research, and clinical applications \cite{OpenAnatomyWeb}. We believe future applications may create Atlas Information Systems (AIS) as an analogue to GIS systems, with the atlas serving as a visual interface to access data associated with the various segmented objects, with further support for geometric, topological, and ontological queries.

In the spirit of reproducible science, we have implemented the algorithm in both 2D and 3D and contributed them to the VTK system \cite{VTK} using the C++ programming language. Open source implementations are available as \texttt{\small vtkSurface\-Nets2D} and \texttt{\small vtkSurface\-Nets3D}. Also available in VTK are the classes \texttt{\small vtkDiscrete\-Marching\-Cubes} and \texttt{\small vtkDiscrete\-Flying\-Edges\-3D} which are used to compare against the Parallel Surface\-Nets algorithm. PSN smoothing is performed by the \texttt{\small vtkConstrained\-Smoothing\-Filter}; Taubin's algorithm \cite{TAUBIN} used with MC and FE comparisons is implemented in the \texttt{\small vtkWindowedSinc\-Poly\-Data\-Filter}.

\section{ACKNOWLEDGEMENTS}
Portions of this work were funded under NIH grants: NIH R01 from the National Institute of Biomedical Imaging and Bioengineering (NIBIB), “Accelerating Community-Driven Medical Innovation with VTK" (R01EB014955), and NIH R01 from NBIB "Stationary Digital Tomosynthesis for Transbronchial Biopsy Guidance" (R01EB028283).

Thanks to Jakob Troidl and Hanspeter Pfister of the Visual Computing Group at Harvard for their help obtaining and processing the connectomics Pinky dataset. Also Jean-Christophe Fillion-Robin and Matt McCormick at Kitware for data wrangling the Allen Mouse Brain data.

\bibliographystyle{abbrv-doi}

\bibliography{references}

\section*{APPENDIX}

To reproduce the results of this paper, three basic steps must be used.
\begin{enumerate}
    \item Download and build a recent version of VTK with the appropriate compilation options. This provides the core algorithms implementations / classes as described previously in the paper.
    \item Copy the SurfaceNets repository which contains the five datasets used in this paper, and test driver programs to execute the testing process.
    \item Build and run the test driver program(s).
\end{enumerate}
Note that all code is written in C++, requiring a compiler supporting C++11 or later. The build processes for both VTK and SurfaceNets is based on CMake, a recent version (3.23 or later) is required. 

The proper VTK version is 9.2.20230420 (i.e., date stamp Apr. 20, 2023) or later. Instructions for building can be found  \href{https://tinyurl.com/mrx4knub}{\textcolor{blue}{here}}. Make sure to build with \texttt{\small CMAKE\-\_BUILD\-\_TYPE = Release} and {\small VTK\_SMP\_IMPLE\-MENTATION\_TYPE~=~TBB} to enable template optimization and parallel threading using TBB. For fastest builds (as VTK is a large system), we recommend that you use the ninja build tool as follows:
\begin{verbatim}
    cmake -GNinja ../VTK
\end{verbatim}
where this command is issued in a build directory adjacent to the VTK source code.

The SurfaceNets testing repository can be copied from \href{https://tinyurl.com/57brseb9}{\textcolor{blue}{here}}. Copy this gitlab repository to obtain the data files and test driver programs. Again, use CMake to build the programs, which will find and link against the VTK build. The test driver programs are simple instantiations of VTK pipelines for reading, processing, displaying, and writing output data. vtkTimerLogs instances are also used to report timings. A README file is available in the SurfaceNets repository which provides further instructions and information.

A special note regarding Frisken's \cite{Frisken22} Multi-Material SurfaceNets algorithm. The author has granted us permission to include the core algorithmic code into the SurfaceNets testing repository. Thus the test driver repository incorprates MM as part of the testing process. However, the author requests that appropriate citation is used if the MM code is used in future publications.

\end{document}